\documentclass[11pt]{article}
\usepackage{amsmath,amssymb,epsfig,bm}

\date{\empty}

\begin{document}

\title{\bf The deceleration parameter in `tilted' universes: generalising the Friedmann background}

\author{Christos G. Tsagas\\ {\small Section of Astrophysics, Astronomy and Mechanics, Department of Physics}\\ {\small Aristotle University of Thessaloniki, Thessaloniki 54124, Greece}\\ {\small \it and}\\ {\small Clare Hall, University of Cambridge, Herschel Road, Cambridge CB3 9AL, UK}}

\maketitle

\begin{abstract}
Large-scale bulk peculiar motions introduce a characteristic length scale, inside which the local kinematics are dominated by peculiar-velocity perturbations rather than by the background Hubble expansion. Regions smaller than the aforementioned critical length, which typically varies between few hundred and several hundred Mpc, can be heavily ``contaminated'' by the observers' relative motion. For example, at the critical length -- hereafter referred to as the ``transition scale'', the sign of the locally measured deceleration parameter can change from positive to negative, while the surrounding universe is still decelerating globally. Overall, distant observers can assign very different values to their local deceleration parameters, entirely because of their relative motion. In practice, this suggests that information selected from regions inside and close to the transition scale hold only locally and they should not be readily extrapolated to the global universe. We show that this principle applies to essentially all Friedmann backgrounds, irrespective of their equation of state and spatial curvature. Put another way, the transition scale and the related effects are generic to linear peculiar-velocity perturbations. This study generalises previous work applied, primarily for reasons of mathematical simplicity, to a perturbed Einstein-de Sitter universe.
\end{abstract}

\section{Introduction}\label{sI}
Observers moving with respect to each other generally have a different understanding of what one may call reality. This is especially true in relativity, where space and time are no longer separate and absolute entities, but interconnected and relevant. More specifically, in Einstein's theory, observers moving relative to each other have their own time and 3-space and they generally disagree on their temporal and spatial measurements. Applied to cosmology, this principle suggests that distant observers may have a different understanding of the universe they live in, simply because they happen to move with respect to each other. If it so happens, it should not come as a complete surprise. After all, in the history of astronomy there are many examples where relative-motion effects have led to a gross misinterpretation of reality.

Real observers in the universe do not simply follow the smooth Hubble expansion, but we all have finite peculiar velocities relative to it. Our Local Group of galaxies, for example, ``drifts'' at a speed of approximately 600~km/sec with respect to the reference frame of the universe~\cite{Ketal,Agetal}. The latter has been traditionally identified with the coordinate system where the dipole of the Cosmic Microwave Background (CMB) vanishes. Analogous peculiar motions, though on much larger scales, have been repeatedly reported by numerous surveys. These are the so-called ``bulk flows'', with typical sizes from few hundred to several hundred Mpc and typical velocities ranging between few hundred and several hundred km/sec (e.g.~see~\cite{Adetal} and references therein). As a result, in addition to their peculiar motion relative to the Hubble  frame, observers in distant galaxies are expected to move with respect to each other as well. All these local deviations from the mean Hubble flow are believed to be a fairly recent (post-recombination) addition to the kinematics of our cosmos, triggered by the ongoing process of structure formation.

Relativistic studies of large-scale peculiar motions require the use of the so-called ``tilted'' cosmological models (e.g.~see~\cite{KE,HUW}). These allow for two (at least) families of observers moving with respect to each other, with a hyperbolic ``tilt'' angle forming between their 4-velocities. Typically, one family is identified with the (idealised) reference frame of the universe, namely with the Hubble/CMB frame, relative to which peculiar velocities can be defined and measured. The second group of observers is assumed to reside in typical galaxies, like our Milky Way, which ``drift'' with respect to the mean universal expansion. Alternatively, one may align the aforementioned two 4-velocity fields with the rest-frames of distant observers, moving relative to each other.

Overall, tilted cosmologies provide a more realistic representation of the current universe, as they facilitate the analysis of multi-component systems (e.g.~\cite{TCM,EMM}). This is probably the reason for the recent increase in the number of these studies. In~\cite{HD-PI}, for example, the reader can find an analysis of tilted Lemaitre-Tolman-Bondi spacetimes. More recently, tilted Friedmann models were employed to investigate the evolution of peculiar-velocity perturbations~\cite{TT}. Peculiar velocities in Szekeres-II regions, matched to a $\Lambda$CDM background, were analysed in~\cite{NS}, as a means to alleviate the Hubble-tension problem. Tilted universes were also recently used to study the implications of dark-matter decay in voids~\cite{LB}.

Employing tilted almost-Friedmann models, the implications of large-scale peculiar motions for the way the bulk-flow observers interpret their cosmological data were studied in~\cite{T1}-\cite{T3}, where we refer the reader for further details and discussion. It was argued there that relative-motion effects can drastically change the local value of the deceleration parameter. More specifically, the local deceleration parameter measured by observers residing in a slightly expanding bulk flow can be considerably larger than that of the actual universe, due to relative-motion effects alone. Observes that happen to live in slightly contracting bulk motions, on the other hand, may assign much smaller values to their local deceleration parameter. Most intriguingly, for some of the latter observers, the sign of the deceleration parameter may also change from positive to negative. In every case, the effect is local caused by the observers' peculiar motion. Nevertheless, the affected scales can be large enough (typically between few hundred and several hundred Mpc) to create the false impression of a recent global event. It is therefore, likely that an unsuspecting bulk-flow observer may misinterpret an apparent local change in the sign of the deceleration parameter as recent global acceleration~\cite{TK,T3}.

Observers inside large-scale bulk flows can be misled to erroneous conclusions, because their observational data are ``contaminated'' by their own peculiar motion. The latter sets a characteristic length scale, within which the relative-motion effects dominate over the background expansion and thus dictate the local linear kinematics. This ``transition scale'', which is closely analogous to the familiar Jeans length, typically varies between few hundred and several hundred Mpc, depending on the speed and the size of the bulk flow~\cite{T3}. The effect is purely relativistic, with no known Newtonian analogue, due to the fundamentally different way the two theories treat the gravitational field and its sources (see \S~\ref{sPRE} below and also~\cite{TKA} for a comparison between the relativistic and the Newtonian studies). Given that no real observer in the universe follows the smooth Hubble expansion, we are all surrounded by a ``transition region'', within which our cosmological data can be heavily contaminated by relative-motion effects. As a result, the local deceleration parameter is larger/smaller than that of the actual universe, depending on whether the associated bulk is locally expanding/contracting. Then, assuming that expanding and contracting bulk flows are evenly distributed, half of the observers in the universe may think that their cosmos is over-decelerated, while the rest may come to believe that they live in an under-decelerated, or even in an accelerated, world~\cite{T3}. In every case, the unsuspecting observers have been misled by their peculiar motion with respect to the mean Hubble flow.

So far, the studies looking into the implications of bulk peculiar motions for the interpretation of the global kinematics of the universe were confined to a perturbed Einstein-de Sitter cosmology, with a 4-velocity tilt~\cite{T1}-\cite{T3}. The latter was selected for reasons of mathematical simplicity and not because the relative-motion effects outlined above were exclusive to the Einstein-de Sitter universe. Here, we will verify this by extending the analysis to tilted Friedmann-Robertson-Walker (FRW) models with varying spatial curvature and equation of state. Among others, these spacetimes include all FRW cosmologies with matter that satisfies the strong energy condition. We will show, in particular, that real observers in these universes are surrounded by an extended region, namely by their own transition domain. There, the relative-motion effects dominate over the background Hubble expansion and, in so doing, they can ``contaminate'' the cosmological data and interfere with the interpretation of the observations. Our analysis also suggests that this is a generic relative-motion effect, independent of the symmetries of the background spacetime. In other words, the formation of a transition scale appears to be as generic to linear peculiar-velocity perturbations, as the Jeans length is to linear density perturbations. This means that local measurements made by observers residing in distant parts of the universe may differ wildly and thus lead them to entirely different conclusions about the kinematics of their cosmos. One should therefore be very careful before extrapolating locally collected data to the global universe.

\section{Aspects of relative motion}\label{sARM}
\begin{figure}[tbp]
\centering \vspace{6cm} \includegraphics{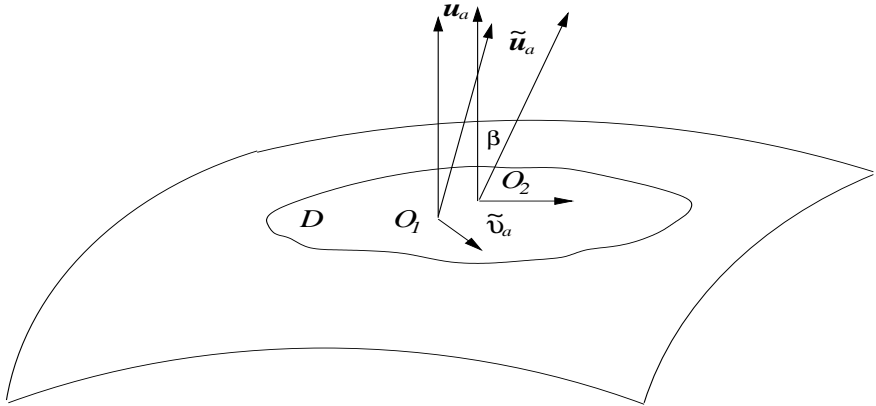} \caption{Observers ($O_1$, $O_2$) inside a bulk-flow ($D$), with peculiar velocity $\tilde{v}_a$ relative to the Hubble expansion.  The 4-velocities $u_a$ and $\tilde{u}_a$, with a hyperbolic ``tilt'' angle ($\beta$) between them, define the reference (Hubble) frame and that of the bulk peculiar motion respectively (see Eq.~(\ref{Lorentz1})). One may also align these two 4-velocity fields with distant observers in the universe that happen to move relative to each other.}  \label{fig:bflow}
\end{figure}

Consider two groups of relatively moving observers with timelike 4-velocities $u_a$ and $\tilde{u}_a$, so that $u_au^a=-1=\tilde{u}_a\tilde{u}^a$. Assuming that $v_a$ is the relative velocity of the latter group with respect to the former, we have
\begin{equation}
\tilde{u}_a= \tilde{\gamma}(u_a+\tilde{v}_a)\,, \label{Lorentz1}
\end{equation}
where $u_a\tilde{v}^a=0$ by construction and $\tilde{\gamma}=(1-\tilde{v}^2)^{-1/2}$ is the Lorentz-boost factor (with $\tilde{v}^2=\tilde{v}_a\tilde{v}^a$). The above transformation also determines the hyperbolic ``tilt'' angle ($\beta$, with $\cosh\beta=- u_a\tilde{u}^a=\tilde{\gamma}>1$) between the two 4-velocity fields (e.g.~see~\cite{KE,HUW} and also Fig.~\ref{fig:bflow} here). Assuming non-relativistic relative motions, with $\tilde{v}^2\ll1$ and $\tilde{\gamma}\simeq1$, the Lorentz boost reduces to
\begin{equation}
\tilde{u}_a= u_a+ \tilde{v}_a\,.  \label{Lorentz2}
\end{equation}
Hereafter, we will treat the $u_a$-field as our reference frame, with respect to which peculiar velocities can be defined and measured. The ``tilted'' (i.e.~the $\tilde{u}_a$) field, on the other hand, will define the rest-frame of any other relatively moving observer. For example, of an observer living inside a large-scale bulk flow, like those reported in many surveys (see Fig.~\ref{fig:bflow}). Alternatively, one could identify these two 4-velocity fields with distant observers in the universe, moving relative to each other.

In accord with the 1+3 covariant approach to relativity and cosmology (see~\cite{Eh,El}, as well as~\cite{TCM,EMM} for recent extensive reviews), the two 4-velocity fields, $u_a$ and $\tilde{u}_a$ set the time direction of the associated observers. At the same time, the projectors $h_{ab}=g_{ab}+u_au_b$ and $\tilde{h}_{ab}=g_{ab}+ \tilde{u}_a\tilde{u}_b$ (with $h_{ab}u^b=0= \tilde{h}_{ab}\tilde{u}^b$ and $g_{ab}$ being the spacetime metric) act as the metric tensors of the 3-D spatial sections. Note that $h_{ab}=h_{(ab)}$, $h_{ab}=h_{ac}h^c{}_b$ and $h_a{}^a=3$ by construction, with analogous relations holding in the tilted frame as well (e.g.~see~\cite{TCM,EMM}). Employing the two 4-velocity fields and their corresponding projection tensors, one defines temporal and spatial derivatives in the two frames. Here, we will use overdots and primes to indicate time differentiation in the $u_a$-frame and in its tilted counterpart respectively (i.e.~${}^{\cdot}=u^a\nabla_a$ and $~{}^{\prime}= \tilde{u}^a\nabla_a$, with $\nabla_a$ being the 4-D covariant derivative). The associated spatial covariant derivative operators, on the other hand, will be  ${\rm D}_a=h_a{}^b\nabla_b$ and $\tilde{{\rm D}}_a= \tilde{h}_a{}^b\nabla_b$ (orthogonal to the $u_a$ and to the $\tilde{u}_a$ field respectively).

Both frames are physically equivalent and one is free to choose either of them to describe the peculiar kinematics, since all that matters is their relative motion. In what follows, we will select the tilted coordinate system and thus adopt the perspective of a real observers, namely those living in a typical galaxy like our Milky Way and drifting with respect to the mean universal expansion. Our frame choice makes it easier to incorporate all the relative-motion effects into the analysis, while it provides a direct and transparent interpretation of the results as well (see also \S~\ref{sCFs} below). Besides, the tilted frame is also the coordinate system where the observations take place.

\section{Linear relations between the two frames}\label{sLRBTFs}
All the information determining the kinematics of the two aforementioned frames is encoded in the gradients of their 4-velocity fields, namely in $\nabla_bu_a$ and $\nabla_b\tilde{u}_a$. We may decode this information by introducing the decomposition
\begin{equation}
\nabla_bu_a= {1\over3}\,\Theta h_{ab}+ \sigma_{ab}+ \omega_{ab}- A_au_b\,,  \label{Nbua}
\end{equation}
applied here to the $u_a$-field. The volume scalar $\Theta={\rm D}^au_a$ indicates expansion/contraction when positive/negative, while it is also related to the Hubble parameter by means of $H=\Theta/3$. The symmetric and trace-free shear tensor $\sigma_{ab}={\rm D}_{\langle b}u_{a\rangle}$ monitors shape distortions, the antisymmetric vorticity tensor $\omega_{ab}={\rm D}_{[b}u_{a]}$ implies rotation and the 4-acceleration vector $A_a=\dot{u}_a= u^b\nabla_bu_a$ ensures that non-gravitational forces are in action and vanishes when the observers' worldlines are timelike geodesics. It goes without saying that the covariant derivative of the $\tilde{u}_a$-field splits in an exactly analogous manner, that is
\begin{equation}
\nabla_b\tilde{u}_a= {1\over3}\,\tilde{\Theta}\tilde{h}_{ab}+ \tilde{\sigma}_{ab}+ \tilde{\omega}_{ab}- \tilde{A}_a\tilde{u}_b\,,  \label{tNbua}
\end{equation}
where now $\tilde{\Theta}=\tilde{\rm D}^a\tilde{u}_a$, $\tilde{\sigma}_{ab}= \tilde{{\rm D}}_{\langle b}\tilde{u}_{a\rangle}$, $\tilde{\omega}_{ab}=\tilde{{\rm D}}_{[b}\tilde{u}_{a]}$ and $\tilde{A}_a=\tilde{u}^{\prime}_a= \tilde{u}^b\nabla_b\tilde{u}_a$. Similarly, expressed in the tilted coordinate system,  the gradient of the peculiar velocity field
decomposes as~\cite{ET}
\begin{equation}
\tilde{\rm D}_b\tilde{v}_a= {1\over3}\,\tilde{\vartheta}\tilde{h}_{ab}+ \tilde{\varsigma}_{ab}+ \tilde{\varpi}_{ab}\,,  \label{tDbtva}
\end{equation}
with $\tilde{\vartheta}=\tilde{\rm D}^a\tilde{v}_a$, $\tilde{\varsigma}_{ab}= \tilde{{\rm D}}_{\langle b}\tilde{v}_{a\rangle}$ and $\tilde{\varpi}_{ab}=\tilde{{\rm D}}_{[b}\tilde{v}_{a]}$ representing the expansion/contraction, the shear and the vorticity of the peculiar motion. When the latter is non-relativistic and the background universe is an FRW model, the three velocitiy fields introduced above satisfy the linear relations
\begin{equation}
\tilde{\Theta}= \Theta+ \tilde{\vartheta}\,, \hspace{15mm} \tilde{\sigma}_{ab}= \sigma_{ab}+ \tilde{\varsigma}_{ab}  \label{lrels1a}
\end{equation}
and
\begin{equation}
\tilde{\omega}_{ab}= \omega_{ab}+ \tilde{\varpi}_{ab}\,, \hspace{15mm} \tilde{A}_a= A_a+ v_a^{\prime}+ Hv_a\,,  \label{lrels1b}
\end{equation}
where $H=\dot{a}/a$ is the background Hubble parameter and $a=a(t)$ is the cosmological scale factor~\cite{M}. Expression (\ref{lrels1a}a) implies that the expansion rates measured in the two frames differ due to relative-motion effects alone.\footnote{Although $\tilde{\Theta}$ and $\Theta$ are always positive, due to the global expansion of the universe, the peculiar volume scalar can be either positive of negative (i.e.~$\tilde{\vartheta}\gtrless0$, with $|\tilde{\vartheta}|/\Theta\ll1$ throughout the linear regime). In the former instance the associated bulk flow (see region $D$ in Fig.~\ref{fig:bflow}) is locally expanding (slightly), while in the latter it is locally contracting.} The same is also true for the shear, the vorticity and the 4-acceleration.\footnote{In general, both frames are allowed to have nonzero shear and vorticity at the linear level. Nevertheless, neither of these variables is involved in our calculations, which makes them irrelevant for the rest of the analysis.} As far as the latter is concerned, Eq.~(\ref{lrels1b}b) ensures that there is always an effective 4-acceleration triggered by the observers' peculiar flow (i.e.~$\tilde{A}_a=\tilde{v}_a^{\prime}+ H\tilde{v}_a\neq0$, even when $A_a=0$ -- and vice versa). Put another way, we cannot simultaneously treat the worldlines of the relatively moving observers as timelike geodesics.

The dynamical variables measured in the two frames also differ due to relative-motion effects. More specifically, with respect to the $u_a$-field, the energy-momentum tensor of the matter decomposes as
\begin{equation}
T_{ab}= \rho u_au_b+ {1\over3}\,p h_{ab}+ 2q_{(a}u_{b)}+ \pi_{ab}\,.  \label{Tab}
\end{equation}
Here, $\rho$, $p$, $q_a$ and $\pi_{ab}$ are respectively the matter density, the (isotropic) pressure, the energy flux and the viscosity ``seen'' by the associated observers. An exactly analogous decomposition also holds in the tilted frame, with the two sets of variables related by~\cite{M}
\begin{equation}
\tilde{\rho}= \rho\,, \hspace{15mm} \tilde{p}= p\,, \hspace{15mm} \tilde{q}_a= q_a- (\rho+p)\tilde{v}_a  \label{lrels2a}
\end{equation}
and
\begin{equation}
\tilde{\pi}_{ab}= \pi_{ab}\,,  \label{lrels2b}
\end{equation}
at the linear perturbative level. According to the above, there is no difference in the density and the pressure (both the isotropic and the anisotropic) between the two frames, but there is a difference in the flux vectors. Expression (\ref{lrels2a}c) implies that, even when one observer ``sees'' the cosmic medium as a perfect fluid, any other relatively moving observer will ``see'' it as imperfect. Put another way, in the presence of relative motions, we cannot set both flux vectors to zero simultaneously. The only exception is on de~Sitter-type inflationary backgrounds (with $p=-\rho$), which are therefore excluded from the rest of our analysis. Note that this does not compromise in any way the aim (as well as the merit) of our study, since we are considering the possibility of accelerated expansion in cosmological models with conventional matter that satisfies the strong energy condition (i.e.~with $\rho+3p>0$).

\section{The deceleration parameters in the two
frames}\label{sDPTFs}
The difference in the expansion rates between the two frames seen in (\ref{lrels1a}a), suggests that the corresponding acceleration/deceleration rates should differ as well. Indeed, taking the time derivative of (\ref{lrels1a}a) relative to the $\tilde{u}_a$-frame and keeping up to linear-order terms, we arrive at
\begin{equation}
\tilde{\Theta}^{\prime}= \dot{\Theta}+ \tilde{\vartheta}^{\prime}\,,  \label{dotThetas1}
\end{equation}
where $\tilde{\Theta}^{\prime}=\tilde{u}^a\nabla_a\tilde{\Theta}$, $\dot{\Theta}=u^a\nabla_a\Theta$ and $\tilde{\vartheta}^{\prime}= \tilde{u}^a\nabla_a\tilde{\vartheta}$~\cite{T1,T2}. Accordingly, $\tilde{\Theta}^{\prime}\neq\dot{\Theta}$ and their difference depends on both the sign and the magnitude of $\tilde{\vartheta}^{\prime}$. Note that, although $\tilde{\vartheta}/\Theta\ll1$ throughout the linear regime, this does not necessarily apply to the ratio of their time derivatives.\footnote{Clearly, if the $\tilde{\vartheta}^{\prime}/\dot{\Theta}$-ratio remains large for a prolonged (in cosmological terms) period, the linear constraint $\tilde{\vartheta}/\Theta\ll1$ will be eventually violated.}

Different values for $\tilde{\Theta}^{\prime}$ and $\dot{\Theta}$ between the two frames imply that the associated deceleration parameters differ as well. Indeed, by definition we have
\begin{equation}
\tilde{q}= -\left(1+{3\tilde{\Theta}^{\prime}\over\tilde{\Theta}^2}\right) \hspace{15mm} {\rm and} \hspace{15mm} q= -\left(1+{3\dot{\Theta}\over\Theta^2}\right)\,, \label{qs1}
\end{equation}
in the tilted and the reference frame respectively. These definitions combine with Eqs.~(\ref{lrels1a}a) and (\ref{dotThetas1}) to give the following linear relation between the deceleration parameters measured in the two frames, namely
\begin{equation}
\tilde{q}= q+ {\tilde{\vartheta}^{\prime}\over6\dot{H}} \left[2+(1+3w)\Omega\right]\,,  \label{tq1}
\end{equation}
with $\Omega=\rho/3H^2$ representing the background density parameter. Therefore, in the absence of peculiar motions (i.e.~when $\tilde{\vartheta}=0=\tilde{\vartheta}^{\prime}$), the two deceleration parameters coincide (as expected). Generally, however, $\tilde{q}\neq q$ and their difference is mainly decided by $\tilde{\vartheta}^{\prime}$, namely by the time evolution of $\tilde{\vartheta}$. Next, we will employ linear relativistic cosmological perturbation theory to derive the required expression, assuming a general FRW background with varying spatial curvature and equation of state. In addition, given that peculiar velocities vanish by default in all FRW backgrounds, our analysis is gauge-invariant as well~\cite{TCM,EMM}.

\section{The peculiar Raychaudhuri equation}\label{sPRE} 
The time evolution of the peculiar volume scalar ($\tilde{\vartheta}$) is determined by the associated Raychaudhuri equation. The latter monitors changes in the mean separation between the flow-lines of the bulk peculiar motion (see Fig.~\ref{fig:bflow} in \S~\ref{sARM} earlier) and it can be obtained by applying the Ricci identities to the peculiar velocity field. On an FRW background, the calculation gives
\begin{equation}
\tilde{\vartheta}^{\prime}= -{1\over3}\,\Theta\tilde{\vartheta}+ \tilde{\rm D}^av_a^{\prime}\,,  \label{pRay1}
\end{equation}
to first approximation~\cite{ET}. Therefore, the time evolution of the peculiar volume scalar is determined by $\tilde{\rm D}^av_a^{\prime}$, that is by the spatial divergence of the time derivative of the peculiar velocity.

Despite the fully relativistic nature of our analysis so far, one can arrive at the same linear expressions using Newtonian physics, with the exception of the linear relation (\ref{lrels2a}c) between the two energy-flux vectors~\cite{TKA}. There are differences in many of the definitions, of course, but for all practical purposes the relations are formally identical. The relativistic and the Newtonian approaches start to diverge when gravity comes into play and the reason is the fundamentally different way these theories treat the gravitational field and its sources. In Newtonian physics gravity is a force, triggered by gradients in the gravitational field, to which only the density of the matter contributes. In Einstein's theory, on the other hand, gravity is the manifestation of spacetime curvature. Moreover, it is not only the matter density that contributes to the energy-momentum tensor and therefore to the local gravitational field.  The pressure, both the isotropic and the anisotropic, as well as the energy flux have their own input too. The flux contribution, in particular, is crucial when dealing with bulk peculiar flows, since the latter introduce a nonzero effective flux to the local energy-momentum tensor (see Eq.~(\ref{lrels2a}c) in~\S~\ref{sLRBTFs} previously). Put another way, in relativity bulk flows gravitate, while in Newtonian physics they do not (see~also~\cite{TKA} for further discussion).

The extra flux contribution to the relativistic gravitational field feeds into the associated conservation laws and eventually emerges in the evolution and the constraint equations of the theory. For our purposes, the key formula is the one monitoring the propagation of density inhomogeneities (see expressions (2.3.1) and (10.101) in~\cite{TCM} and~\cite{EMM} respectively). Linearised in the tilted frame on a general FRW background the latter reads
\begin{equation}
\tilde{\Delta}_a^{\prime}= 3wH\tilde{\Delta}_a- (1+w)\tilde{\mathcal{Z}}_a+ {3aH\over\rho}\, \left(\tilde{q}_a^{\prime}+4H\tilde{q}_a\right)- {a\over\rho}\,\tilde{\rm D}_a\tilde{\rm D}^b\tilde{q}_b\,, \label{tDelta1}
\end{equation}
where $\tilde{\Delta}_a=(a/\rho)\tilde{\rm D}_a\rho$ measures spatial variations in the matter distribution relative to the tilted frame and $w=p/\rho$ is the background barotropic index. Also, the spatial gradient $\tilde{\mathcal{Z}}_a=a\tilde{\rm D}_a\tilde{\Theta}$ monitors spatial variations in the universal expansion in the tilted frame as well.

Assuming that the cosmic medium appears perfect in the reference $u_a$-frame, we may set $q_a=0$ there. Then, recalling that $\tilde{q}_a=-(\rho+p)\tilde{v}_a$ in the tilted coordinate system  (see (\ref{lrels2a}c) above), Eq.~(\ref{tDelta1}) recasts into
\begin{equation}
\tilde{\Delta}_a^{\prime}= 3wH\tilde{\Delta}_a -(1+w)\tilde{\mathcal{Z}}_a+ a(1+w)\tilde{\rm D}_a\tilde{\vartheta}- 3a(1+w)H \left[\tilde{v}_a^{\prime}+(1-3c_s^2)H\tilde{v}_a\right]\,,  \label{tDelta2}
\end{equation}
given that $\dot{\rho}=-3H(1+w)\rho$ in the FRW background and with $c_s^2=\dot{p}/\dot{\rho}$ representing the square of the sound speed. Solving the above for $\tilde{v}_a^{\prime}$, taking the spatial divergence of the resulting expression and using the linear commutation law $\tilde{\rm D}^a\tilde{v}_a^{\prime}=(\tilde{\rm D}^a\tilde{v}_a)^{\prime}+ H\tilde{\rm D}^a\tilde{v}_a$, we arrive at
\begin{equation}
\tilde{\vartheta}^{\prime}= -2\left(1-{3\over2}\,c_s^2\right)H\tilde{\vartheta}+ {1\over3H}\, \tilde{\rm D}^2\tilde{\vartheta}- {1\over3a^2(1+w)} \left[{\tilde{\Delta}^{\prime}\over H}-3w\tilde{\Delta} +(1+w){\tilde{\mathcal{Z}}\over H}\right]\,,  \label{pRay2}
\end{equation}
with $\tilde{\Delta}=a\tilde{\rm D}^a\tilde{\Delta}_a$ (with $\tilde{\Delta}^{\prime}=a\tilde{\rm D}^a\tilde{\Delta}_a^{\prime}$ to first approximation) and $\tilde{\mathcal{Z}}=a\tilde{\rm D}^a\tilde{\mathcal{Z}}_a$.\footnote{The scalar $\Delta$ closely corresponds to the familiar density contrast $\delta=\delta\rho/\rho$ of the non-covariant studies~\cite{TCM,EMM}.} The above is the Raychaudhuri equation of the peculiar motion, linearised around an FRW background of arbitrary curvature and equation of state (with the exception of the $w=-1$ case).

\section{Relative motion effects on $\tilde{q}$}\label{sR-MEtq} 
Expression (\ref{pRay2}) determines the linear relative-motion effects on the deceleration parameter (see Eq.~(\ref{tq1}) in \S~\ref{sDPTFs}). To proceed further, we recall that the background Raychaudhuri equation takes the form $\dot{H}=-H^2[1+[(1+3w)\Omega/2]]$, which combined with the above leads to
\begin{eqnarray}
{\tilde{\vartheta}^{\prime}\over6\dot{H}} \left[2+(1+3w)\Omega\right]&=& {2\over3}\left(1-{3\over2}\,c_s^2\right){\tilde{\vartheta}\over H}- {1\over9H^3}\,\tilde{\rm D}^2\tilde{\vartheta} \nonumber\\ &&+{1\over9a^2(1+w)H^2} \left[{\tilde{\Delta}^{\prime}\over H} -3w\tilde{\Delta}+(1+w){\tilde{\mathcal{Z}}\over H}\right]\,.  \label{pRay3}
\end{eqnarray}
This result provides the ``correction term'' that separates the two deceleration parameters in Eq.~(\ref{tq1}). Before, substituting (\ref{pRay3}) into the right-hand side of (\ref{tq1}), however, it is worth noting the spatial Laplacian on the right-hand side of the above. This term introduces a crucial scale dependance, which becomes explicit after a simple harmonic decomposition. Indeed, splitting the perturbed variables harmonically, expression (\ref{pRay3}) gives
\begin{eqnarray}
{\tilde{\vartheta}_{(n)}^{\prime}\over6\dot{H}} \left[2+(1+3w)\Omega\right]&=& {2\over3}\left[1-{3\over2}\,c_s^2 +{1\over6}\left({\lambda_H\over\lambda_n}\right)^2\right] {\tilde{\vartheta}_{(n)}\over H} \nonumber\\ &&+{1\over9(1+w)}\left({\lambda_H\over\lambda_K}\right)^2 \left[{\tilde{\Delta}_{(n)}^{\prime}\over H} -3w\tilde{\Delta}_{(n)}+(1+w){\tilde{\mathcal{Z}}_{(n)}\over H}\right]\,.  \label{pRay4}
\end{eqnarray}
for the $n$-th harmonic mode.\footnote{In deriving (\ref{pRay4}) we have set $\tilde{\vartheta}= \Sigma_n\tilde{\vartheta}_{(n)}\mathcal{Q}^{(n)}$, $\tilde{\Delta}= \Sigma_n\tilde{\Delta}_{(n)}\mathcal{Q}^{(n)}$ and $\tilde{Z}= \Sigma_n\tilde{Z}_{(n)}\mathcal{Q}^{(n)}$, where $\tilde{\rm D}_a\tilde{\vartheta}_{(n)}=0=\tilde{\rm D}_a \tilde{\Delta}_{(n)}=\tilde{\rm D}_a\tilde{Z}_{(n)}$ and $n$ represents the comoving eigenvalue of the harmonic mode. Also, $\mathcal{Q}^{(n)}$ are standard scalar harmonic functions, with $\mathcal{Q}^{\prime(n)}=0$ and $\tilde{\rm D}^2\mathcal{Q}^{(n)}=-(n/a)^2\mathcal{Q}^{(n)}$.} Note that $\lambda_H=1/H$ is the Hubble horizon, $\lambda_n=a/n$ is the physical wavelength of the peculiar-velocity perturbation and $\lambda_K=a/|K|$ (with $K=\pm1$) is the curvature scale of the universe.

Finally, replacing the correction term on the right-hand side of (\ref{tq1}) with (\ref{pRay4}) and dropping the mode-index ($n$) for the economy of the presentation, we arrive at
\begin{equation}
\tilde{q}= q+ {2\over3}\left[1-{3\over2}\,c_s^2 +{1\over6}\left({\lambda_H\over\lambda}\right)^2\right] {\tilde{\vartheta}\over H}+ {|1-\Omega|\over9(1+w)} \left[{\tilde{\Delta}^{\prime}\over H} -3w\tilde{\Delta}+(1+w){\tilde{\mathcal{Z}}\over H}\right]\,,  \label{tq2}
\end{equation}
since $(\lambda_H/\lambda_K)^2=|1-\Omega|$ to zero order. The above holds on all FRW backgrounds, irrespective of the their spatial curvature and equation of state (with the exception of the $w=-1$ models -- see relation (\ref{lrels2a}c) in \S~\ref{sLRBTFs} and discussion there in). For instance, on the Einstein-de Sitter background where $w=0=c_s$ and $\Omega=1$, expression (\ref{tq2}) reduces to the one derived in~\cite{TK,T3}. Clearly, on subhorizon scales -- where $\lambda_H/\lambda\gg1$, the dominant (correction) term on the right-hand side of the above is the one with the aforementioned ratio. The rest have coefficients of order unity or less, given that $w\neq-1$ and provided the density parameter ($\Omega$) does not take unrealistically large values.\footnote{Given that the perturbations $\tilde{\vartheta}/H$, $\Delta$ and $\tilde{\mathcal{Z}}/H$ are all very small during the linear regime, Eq.~(\ref{tq2}) suggests that $\tilde{q}\rightarrow q$ on scales close and beyond the Hubble horizon.} On these grounds, well inside the Hubble scale, the two deceleration parameters are related by
\begin{equation}
\tilde{q}= q+ {1\over9}\left({\lambda_H\over\lambda}\right)^2 {\tilde{\vartheta}\over H}\,,  \label{tq3}
\end{equation}
with $\tilde{\vartheta}/H\ll1$ at the linear level. The above coincides with the linear relation obtained on an Einstein-de Sitter background~\cite{TK,T3}. Therefore, on subhorizon scales, the linear relative-motion effects on the deceleration parameter are the same irrespective of the FRW background.

Expression (\ref{tq3}) provides the deceleration parameter measured inside bulk flows evolving on a general FRW background. In particular, the second term on the right-hand side of the above is the linear relativistic correction, triggered by the observer's peculiar motion. Following (\ref{tq3}), the local value of the deceleration parameter ($\tilde{q}$) is more sensitive to the scale-ratio $\lambda_H/\lambda$ (which results from the 3-D Laplacian seen in Eq.~(\ref{pRay3}) above) than to the kinematic-ratio $\tilde{\vartheta}/H$. This implies that $\tilde{q}$ and $q$ can differ substantially, even during the linear regime -- throughout which $\tilde{\vartheta}/H\ll1$, on scales well inside the Hubble horizon, namely those with $\lambda_H/\lambda\gg1$.

Inside (slightly) expanding bulk flows, in particular, the locally measured deceleration parameter can take values considerably larger than its global counterpart. Observers residing in slightly contracting bulk flows, on the other hand, will assign smaller values to their local deceleration parameter, while some of them may even ``see'' the sign of $\tilde{q}$ to change from positive to negative. Assuming that expanding and contracting bulk flows are randomly distributed, nearly half of the observers in the universe may think that their cosmos is over-decelerated and the other half that it is under-decelerated, or even accelerated. In every case, however, the observers are experiencing an illusion triggered by local relative-motion effects. A Table with representative values for $\tilde{q}$, based on the peculiar velocities reported by a number of recent surveys (e.g.~\cite{FWH}-\cite{Scetal}), has been given in~\cite{T3}, where we refer the reader for more details. In what follows, we will estimate the scale where the correction term on the right-hand side of Eq.~(\ref{tq3}) starts becoming important.

\section{The transition scale}\label{sTS}

\begin{figure}[tbp]
\centering \vspace{4cm} \includegraphics{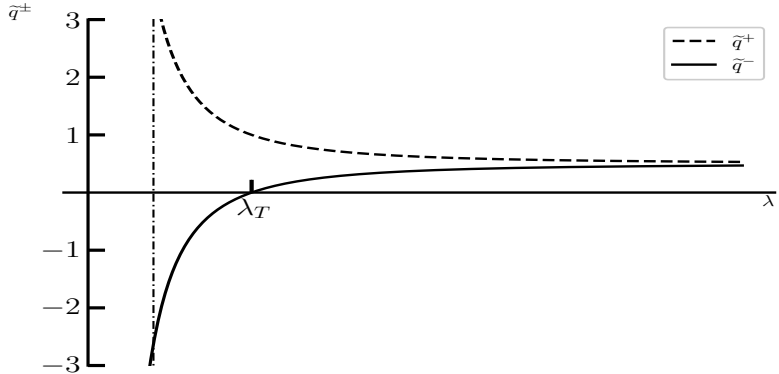} \caption{The transition scale $\lambda_T$ applied here to an Einstein-de Sitter background with $q=1/2$. On scales much larger than $\lambda_T$ the relative-motion effects are negligible and $\tilde{q}^{\,\pm}\rightarrow1/2$. Scales close and inside the transition scale, on the other hand, are heavily contaminated by the observers peculiar motion. There, $\tilde{q}^{\,+}$ becomes increasingly more positive (dashed curve), while $\tilde{q}^{\,-}$ turns negative at $\lambda_T$ and keeps decreasing on progressively smaller wavelengths (solid curve). The vertical line marks the nonlinear cutoff, where the linear approximation breaks down.}  \label{fig:tq+-}
\end{figure}

The strong scale dependence seen in Eq.~(\ref{tq3}) suggests that on sufficiently small scales the relative-motion effects dominate its right-hand side and thus determine the local value of $\tilde{q}$. This happens when the correction term, due to the tilted observer's peculiar motion, equals (in absolute terms) the value of the deceleration parameter ($q$) measured in the reference $u_a$-frame. It is then straightforward to show that the critical length, where the relative-motion effects start to dominate, is given by
\begin{equation}
\lambda_T= \sqrt{{1\over9q}\,{|\tilde{\vartheta}|\over H}}\, \lambda_H\,,  \label{lambdaT}
\end{equation}
hereafter referred to as the ``transition scale''.\footnote{In~\cite{T3} the transition scale was named the ``peculiar Jeans length'' to emphasise the close analogy between these two critical lengths. Recall that the Jeans length marks the scale below which linear pressure-gradient perturbations dominate over the background gravitational pull and thus dictate the local evolution of linear density perturbations (e.g.~see~\cite{TCM,EMM}). Here, instead, it is the linear peculiar-velocity perturbations that dominate over the background Hubble expansion on scales smaller than $\lambda_T$ (see also~\cite{T3} for further discussion).} Indeed, substituting the above definition back into (\ref{tq3}), the latter recasts into
\begin{equation}
\tilde{q}^{\,\pm}= q\left[1\pm\left({\lambda_T\over\lambda}\right)^2\right]\,,  \label{tq4}
\end{equation}
where the $+/-$ sign implies locally expanding/contracting bulk flows. According to the above, on scales larger than $\lambda_T$ the two deceleration parameters essentially coincide (i.e.~$\tilde{q}\rightarrow q$). On wavelengths smaller than the transition scale, however, the deceleration parameter measured by the tilted observers can differ considerably from the one measured by those in the reference frame. Indeed, following (\ref{tq4}) we find that $\tilde{q}^{\,+}>2q$ and $\tilde{q}^{\,-}<0$ when $\lambda<\lambda_T$ (see~\cite{T3} and also Fig.~\ref{fig:tq+-} here). Clearly, on wavelengths much smaller than the transition scale, the difference between $\tilde{q}^{\pm}$ and $q$ increases further. One should always keep in mind of course that below a certain threshold, which is typically assumed to be around 100~Mpc, the nonlinear effects are no longer negligible and our linear approximation breaks down (see vertical line in Fig.~\ref{fig:tq+-}).

\section{On the choice of frames}\label{sCFs}
In our analysis so far, we have assumed that there is a preferred reference frame in the universe, relative to which one can define and measure peculiar velocities. This coordinate system, which is selected naturally by the universal expansion, is referred to as the Hubble-frame and it is usually identified with the coordinate system where the dipole in the CMB vanishes. Here, the $u_a$-field corresponds to such an idealised reference frame, while the tilted 4-velocity field is associated with the real observers that move relative to the Hubble flow. This relative motion is responsible for contaminating the cosmological data (within and near the transition region of radius $\lambda_T$) and for misdirecting the real observers to erroneous conclusions about the global kinematics of their host universe.

Here, as well as in~\cite{T1}-\cite{T3}, we have taken the perspective of the tilted (the real) observers, namely those living in typical galaxies (like our Milky Way) that move relative to the Hubble flow. This choice made it easier to locate and incorporate the crucial contribution of the bulk-flow flux to the local gravitational field, which explains why the aforementioned flux input was overlooked/bypassed in earlier relativistic studies that took the perspective of the (fictitious) Hubble-flow observers. An additional reason for adopting the viewpoint of the real observers is that their rest-frame is also the coordinate system where the observations take place. Physics is not frame-dependent, of course. Therefore, it is not the frame choice that matters the most, but their relative motion. Indeed, the same results can also be reached by adopting the perspective of the Hubble-flow observers. To demonstrate this, one simply needs to account for the fact that the Hubble frame has ``peculiar'' velocity $v_a=-\tilde{v}_a$ relative to the tilted frame (at the linear level). This in turn ensures that there is an associated energy flux ($q_a=-\rho v_a$), which contributes to the local energy-momentum tensor. By so doing, one arrives at the same conclusions with those reached in \S~\ref{sR-MEtq} and \S~\ref{sTS} here (see~\cite{T3,TKA} for the details). Overall, although changing the observer's perspective is physically irrelevant, as it should be, the tilted frame seems better suited to the study peculiar velocities in cosmology.

The presence a preferred coordinate system in the universe, makes it meaningful to define, measure and study peculiar motions in cosmology. Nevertheless, the assumption of the Hubble/CMB-frame is not essential for the generality of our results. After all, there is always the chance that such a preferred coordinate system may not even exist (see \S~\ref{sD} next). Of course, without accounting for such a global reference frame, it is meaningless to talk about peculiar velocities and all that matters is the relative motion between any two distant observers in the universe. Nevertheless, even in this case our results still hold, although now both 4-velocity fields correspond to real observers, living in distant galaxies and moving relative to each other. It makes no difference which observer's rest-frame is treated as the reference coordinate system and which one is identified with the tilted frame, since all that matters is their relative motion. The latter guarantees that there still exist a transition region, within and near which the values of the deceleration parameter measured by the aforementioned observers can vary wildly. By and large, the formation of the transition scale appears to be generic to bulk peculiar motions, irrespective of the specifics of the host universe.

\section{Discussion}\label{sD}
Recent observations have triggered discussions within the cosmological community, which could potentially have far reaching implications. The catalyst for the ongoing debate appears to be so far unexplained differences between the early and the late-time measurements of the Hubble constant~\cite{Retal}. As a result, an increasing number of cosmologists are reconsidering the current cosmological paradigm and even question the FRW models and the Cosmological Principle itself~\cite{Ketal1}-\cite{Letal}. Persistent reports of anomalous dipolar anisotropies in the number counts of distant radio-sources have added to this debate~\cite{S3}-\cite{NDKP}. In summary, it is conceivable that the time may have come to review some of our long-standing views and perceptions regarding the universe we live in.

A widespread perception among the cosmological community is that accelerated expansion is impossible in perturbed Friedmann universes, unless one allows for dark energy or for a positive cosmological constant. This belief has been based on theoretical studies of single-fluid cosmologies, allowing for one family of observers moving along with the mean universal expansion. No real observer follows the Hubble flow, however, but we all have some finite peculiar velocity with respect to it. In practice, this means that one needs to employ tilted cosmological models, equipped with two groups of observers moving relative to each other. Then, even at the linear perturbative level, the standard picture can change drastically. In fact, peculiar motions can lead to apparent acceleration, which although local it affects scales large enough to be of cosmological relevance. Moreover, this is achieved without the need of appealing to exotic forms of matter, modifying general relativity, or abandoning the Friedmann models.

Introducing a second family of observers brings about linear relative-motion effects that were completely unaccounted for in all previous single-fluid studies. More specifically, relative motion induces an effective energy-flux vector, even when the cosmic fluid is perfect, which then contributes to the local gravitational field. Applied to cosmology, this additional flux-input implies that bulk peculiar motions (in a sense) gravitate. Clearly, this is a purely relativistic effect that the Newtonian studies cannot reproduce naturally. As a result, the relativistic formulae for the linear bulk-flow kinematics acquire extra terms/effects due to the observers' peculiar motion. These equations ensure the formation of a domain around the tilted, that is the relatively moving, observers. This is the transition region, where the kinematics are no longer dominated by the background expansion but by relative-motion effects.

Assuming an Einstein-de Sitter background, primarily for reasons of simplicity, it was shown that the value, as well as the sign, of the deceleration parameter can change inside the transition domain~\cite{T1}-\cite{T3}. The local sign-change, in particular, happened when the observers' bulk flow was (slightly) contracting relative to the background expansion (see solid curve in Fig.~\ref{fig:tq+-}). The affected scale, namely the typical size of the transition region, was found to range between few hundred and several hundred Mpc, which made it large enough to create the false impression of global acceleration. Here, we have extended, the work of~\cite{T1}-\cite{T3} to FRW backgrounds with varying spatial curvature and equation of state, including all those that satisfy the strong energy condition. We found that the relative motion of the tilted observers still forms a transition domain around them, where both the local value and even the sign of the deceleration parameter can change. In fact, the formation of the transition region appears to be a generic feature of relative motion, independent of the background symmetries. Here, for instance, the background was Friedmann-like and therefore the transition domain was (nearly) spherically symmetric. We expect that changing to, say, an axially symmetric Bianchi background, should only change the shape of the transition region to reflect the anisotropy of the host spacetime. Overall, the transition length seems as generic to linear peculiar-velocity perturbations, as the Jeans length is to linear density perturbations.

Typically the transition scale ranges from few hundred to several hundred Mpc, which makes it large enough to be of cosmological relevance and thus to create the false impression of a global event. In particular, measuring larger/smaller values for the local deceleration parameter, may give the false impression that the host universe is over/under-decelerated globally. More dramatically, a local change in the sign of their deceleration parameter, may mislead the observers to believe that the whole universe has recently entered a phase of accelerated expansion. If it so happens, these unsuspecting observers would have misinterpreted a local relative-motion effect as a recent global event, namely the apparent change in the sign of the local deceleration parameter as global acceleration.

Provided that locally expanding and contracting bulk flows are evenly distributed, nearly half of the observers in the universe will believe that they live in an over-decelerated cosmos and the rest in an under-decelerated. In fact, some of observers in the latter group may even think that their universe has recently started to accelerate. Nevertheless, there are still ways for these unsuspecting observers to find out that they have been merely experiencing an illusion. The answer must be in the data and the first sign is the redshift distribution of the measured deceleration parameter, which should take negative values at relatively low redsifts and turn positive at higher redshits. In other words, the deceleration parameter should follow the profile of the solid curve seen in Fig.~\ref{fig:tq+-}, which largely agrees with the observations. In fact, a more refined analysis shows that the tilted scenario fits the data as well as the $\Lambda$CDM paradigm~\cite{AKPT}. The observers should also look for the ``trademark signature'' of relative motion, namely for an apparent (Doppler-like) dipolar anisotropy in the sky-distribution of the deceleration parameter~\cite{T2}. The latter should take more negative values towards a certain direction in the sky and equally less negative towards the antipodal. Put another way, to these observers, the universe should appear to accelerate faster along a point on the celestial sphere and equally slower along its antidiametric. Moreover, the axis of the dipole should not lie far away from that in the CMB spectrum, assuming that both are due to the observers' peculiar motion.\footnote{At the same time, one would like to have more and better (more refined) bulk-flow data. It would be very helpful, in particular, to know the peculiar-velocity profile within the bulk flow, rather than the mean bulk velocity only. In that case, one should be able to use Eqs.~(\ref{lambdaT}) and (\ref{tq4}) to reconstruct the predicted profile of the deceleration parameter, as measured by observers living inside the bulk flow in question.}

Over the last decade there have been reports that such a dipole may exist in the data~\cite{CL-B}-\cite{BBA}. Put another way, our universe may actually accelerate faster in one direction and equally slower in the opposite. It was only recently, however, that the reported $q$-dipole was attributed to our peculiar motion, a step that also reduced the statistical significance of the $q$-monopole~\cite{CMRS2}. Although the issue remains open, additional support may come from independent surveys reporting a dipole in the sky-distribution of the Hubble parameter as well~\cite{Metal1,Metal2}. Indeed, given the close relation between the Hubble and the deceleration parameters, a dipolar anisotropy in one should almost inevitably ensure the same for the other.\\

\textbf{Acknowledgements:} The author wishes to thank Kerkyra Asvesta for helpful comments. This work was supported by the Hellenic Foundation for Research and Innovation (H.F.R.I.), under the ``First Call for H.F.R.I. Research Projects to support Faculty members and Researchers and the procurement of high-cost research equipment Grant'' (Project Number: 789).\\

\end{document}